\documentclass[twocolumn,preprintnumbers,amsmath,amssymb]{revtex4}
\usepackage{tabularx,graphicx}

\usepackage{color}
\usepackage{hyperref}
\hypersetup{
    colorlinks=true,
    linkcolor=blue,
    filecolor=blue,      
    urlcolor=blue,
}

\usepackage{color}

\usepackage{ulem}   % to strike things out

\begin{document}
%\documentstyle[aps]{revtex}
%\documentstyle[preprint,aps]{revtex}
%\begin{document}

\newcommand{\beq}{\begin{equation}}
\newcommand{\eeq}{\end{equation}}
\newcommand{\beqn}{\begin{eqnarray}}
\newcommand{\eeqn}{\end{eqnarray}}
\newcommand{\bmath}{\begin{subequations}}
\newcommand{\emath}{\end{subequations}}
\newcommand{\bra}[1]{\langle #1|}
\newcommand{\ket}[1]{|#1\rangle}

%\draft
\title{On magnetic field screening and trapping  in hydrogen-rich high-temperature superconductors: unpulling the wool over readers' eyes}

\author{J. E. Hirsch$^{a}$  and F. Marsiglio$^{b}$ }
\address{$^{a}$Department of Physics, University of California, San Diego,
La Jolla, CA 92093-0319\\
$^{b}$Department of Physics, University of Alberta, Edmonton,
Alberta, Canada T6G 2E1}
 
 \begin{abstract} 
In Nat Commun 13, 3194 (2022) \cite{e2021p}, Minkov et al. reported magnetization measurements on hydrides under pressure
that claimed to find   a diamagnetic signal   below a critical temperature demonstrating the existence
of superconductivity. 
Here we present an analysis of raw data recently released \cite{correction} by the authors of \cite{e2021p}  that shows that the measured  data 
do not support their  claim that the samples exhibit  a diamagnetic response indicative of superconductivity. We also point out that
Ref. \cite{e2021p} in its original form omitted essential information that resulted in presentation of a distorted picture of reality,
and that important information on transformations performed on measured data remains undisclosed. Our analysis also calls into question the
conclusions of Minkov et al's trapped flux experiments reported in Nat. Phys. (2023) \cite{etrappedp} as supporting superconductivity in these
materials. This  work together with earlier work implies that there is no magnetic evidence for the existence of
high temperature superconductivity in hydrides under pressure. \end{abstract}
 \maketitle 

       \begin{figure} [t]
 \resizebox{7.5cm}{!}{\includegraphics[width=6cm]{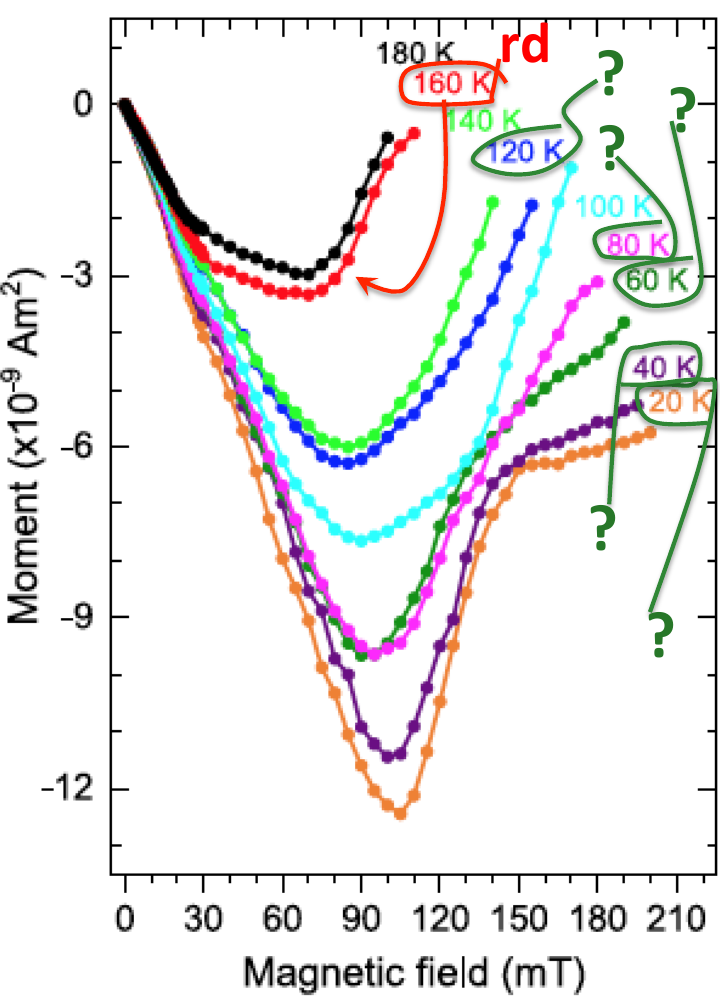}} 
 \caption { Magnetization of $H_3S$ under pressure reported in Ref. \cite{e2021p}. The question marks label curves for which 
 no information exists on what their relation with measured data is. For the single curve labeled ``rd'', some  
 information on underlying raw data has recently been disclosed \cite{correction}.}
 \label{figure1}
 \end{figure}

\section{introduction}

While abundant  evidence from resistance measurements has been put forth claiming that hydrides under high pressure
are high temperature superconductors \cite{review}, magnetic evidence presented  in favor of superconductivity in 
hydrides remains scarce \cite{pers}. Minkov et al. have recently claimed that magnetization measurements \cite{e2021p} 
and trapped flux measurements \cite{etrappedp}  provide clear evidence in favor of superconductivity in these materials.
In this paper we show that their claims are not supported by underlying raw data recently released \cite{correction}. Instead, 
we argue that they are based on the
preconceived assumption that the materials are superconductors  rooted in BCS-Eliashberg theory \cite{pickett,zurek}, and the subsequent interpretation of ambiguous
experimental measurements biased by such prejudice.

Figure 1 shows magnetization measurements reported  by Minkov et al. in Ref. \cite{e2021p}. The caption
of the figure read, when the paper was published on June 9, 2022 and for 450 days thereafter \cite{correction,original}: {\it M(H) magnetization data for Im-3m-H3S   at high pressure. Virgin curves of the M(H) magnetization data for the Im-3m-H3S
phase at PS $= 155 \pm 5$ GPa   at selected temperatures. The curves were superimposed for a better
representation; so the linear trend of M(H) dependences coincides for measurements at different temperatures.''}. The associated
text in the paper read: {\it ``\textbf{M(H) magnetization measurements}. Measurements of the
magnetic field dependence of magnetization allow us to estimate
the characteristic superconducting parameters $H_{c1}, \lambda_L, \kappa$ and $j_c$.  The value of $H_p$, at which the applied magnetic field starts to
penetrate the sample, was determined from the onset of the
deviation of M(H) from the linear dependence (see Fig. 3).''}

These statements informed the reader that the magnetization curves shown in Fig. 1 were measured
in a laboratory.
However, they did not reflect reality. Indeed if the sample showed such behavior, it would be clear evidence that it is diamagnetic, and the magnitude of the signal and the fact that there is a 
clear break for a critical magnetic field suggest that it is superconductivity. However, what is shown in Fig. 1  were not the measured 
data, and this fact was not disclosed to  readers until September 1, 2023 when an ``Author Correction'' was
published \cite{correction}.

The correction \cite{correction} was prompted by emails   from one of us (JEH)  to the authors of Ref. \cite{e2021p} beginning in October 2022 \cite{emails}, asking
the authors to clarify  an  apparent inconsistency between  figures 3a and 3e  of the original version
of Ref. \cite{e2021p}, i.e. Ref. \cite{original}. The authors did not provide
a response to these emails. This was followed by a Matters Arising manuscript by the present authors submitted to Nature Comm in
November 2022 and declined by the journal  in April 2023, that was subsequently published in another
journal, Ref.   \cite{hmscreening}.
In that paper we argued that the  data published in
Ref. \cite{original} were inconsistent with one another and with the expected behavior
of superconductors.

        \begin{figure} [t]
 \resizebox{8.5cm}{!}{\includegraphics[width=6cm]{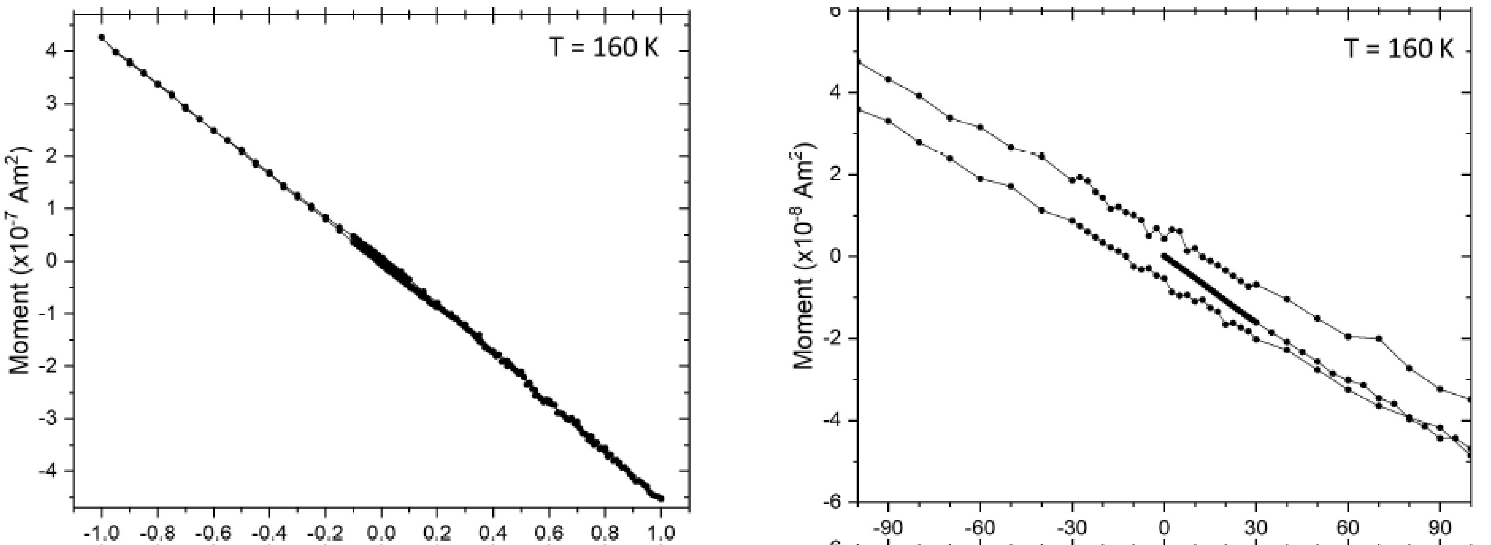}} 
 \caption { From Supplementary Figure S10 of Ref. \cite{e2021p}. The figure caption read:
{\it ``M(H) magnetization data of the heated sample...  The left panel shows the full range of
hysteresis and the right panel shows the enlarged magnetic field range of -0.1 – 0.1 T.''}}
 \label{figure1}
 \end{figure} 
 
         \begin{figure} [t]
 \resizebox{8.5cm}{!}{\includegraphics[width=6cm]{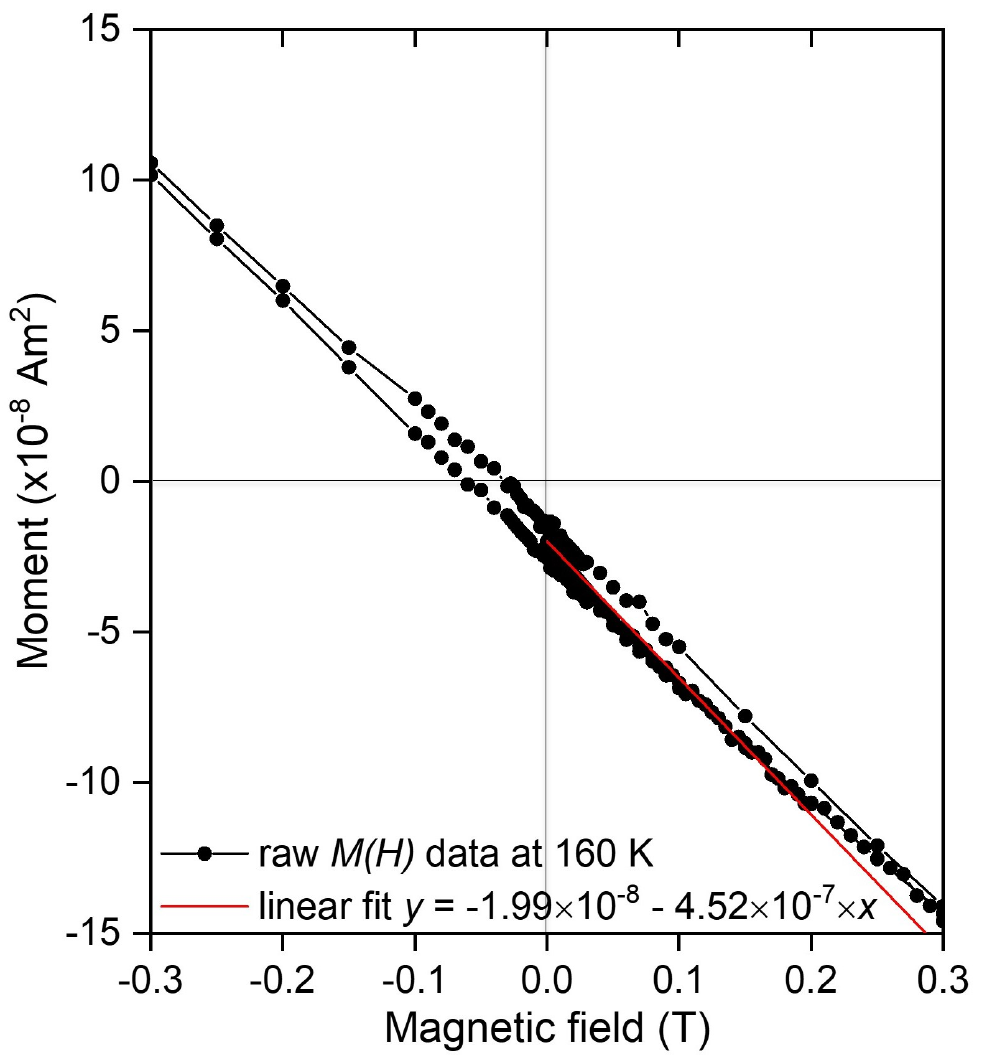}} 
 \caption {Fig. S12b of Ref.  \cite{e2021p}, that was added to the paper when the correction \cite{correction} was published. These are raw data in the range (-0.3T,0.3T), that differ from the ones shown in 
 Fig. 2 by a uniform shift of the data:  the magnetic moment is not zero for zero applied field, as shown by the
 thin horizontal and vertical  lines
 added by us. }
 \label{figure1}
 \end{figure}

The recently published correction \cite{correction} clarifies that the  inconsistency between figures 3a and 3e
of Refs. \cite{e2021p,original}  pointed out in ref. \cite{hmscreening} was
only apparent, not real. The perception of inconsistency originated in the fact that the authors had failed to disclose to readers, and to one of us (JEH)  in
multiple private communications to all the authors where clarification was requested \cite{emails}, that a variety of
transformations had been performed in obtaining the curves shown in Fig. 1 (Fig. 3a of Refs. \cite{e2021p,original}) from
measured data (Fig. 3e of Refs. \cite{e2021p,original}). In this paper we analyze the significance of this disclosure to the
interpretation of the experimental results of Ref. \cite{e2021p} and its implications for Ref. \cite{etrappedp}.

  \section{raw data and transformations for T=160K}
  All the figures in this section refer to the curve labeled ``rd'' in Fig. 1, for T=160K.
The Author Correction Ref. \cite{correction} explains that what was actually measured to infer the curve labeled ``rd'' in Fig. 1 was what is shown in Fig. 2:
a diamagnetic signal, that was non-zero even for zero applied field, that showed some hysteresis when cycling the field between -1T and 1T, starting with the
virgin data  with no field. In fact, Fig. 2 already included a transformation that was not disclosed in the original publication: a vertical shift of the 
data, since the measured signal was  not zero for zero field as Fig. 2 shows, rather it was negative.
Fig. 3 shows the  actual measured data as reported in  ``Supplementary Figure S12b'' of \cite{e2021p}, that 
was  added to the paper when the correction \cite{correction} was published.
% on
%September 1, 2023, as ``Supplementary Figure S12b'', in order to explain 
% the  transformation that was performed to obtain the reported curve labeled 160K in Fig. 1 starting from
%measured data. 

         \begin{figure} [t]
 \resizebox{8.5cm}{!}{\includegraphics[width=6cm]{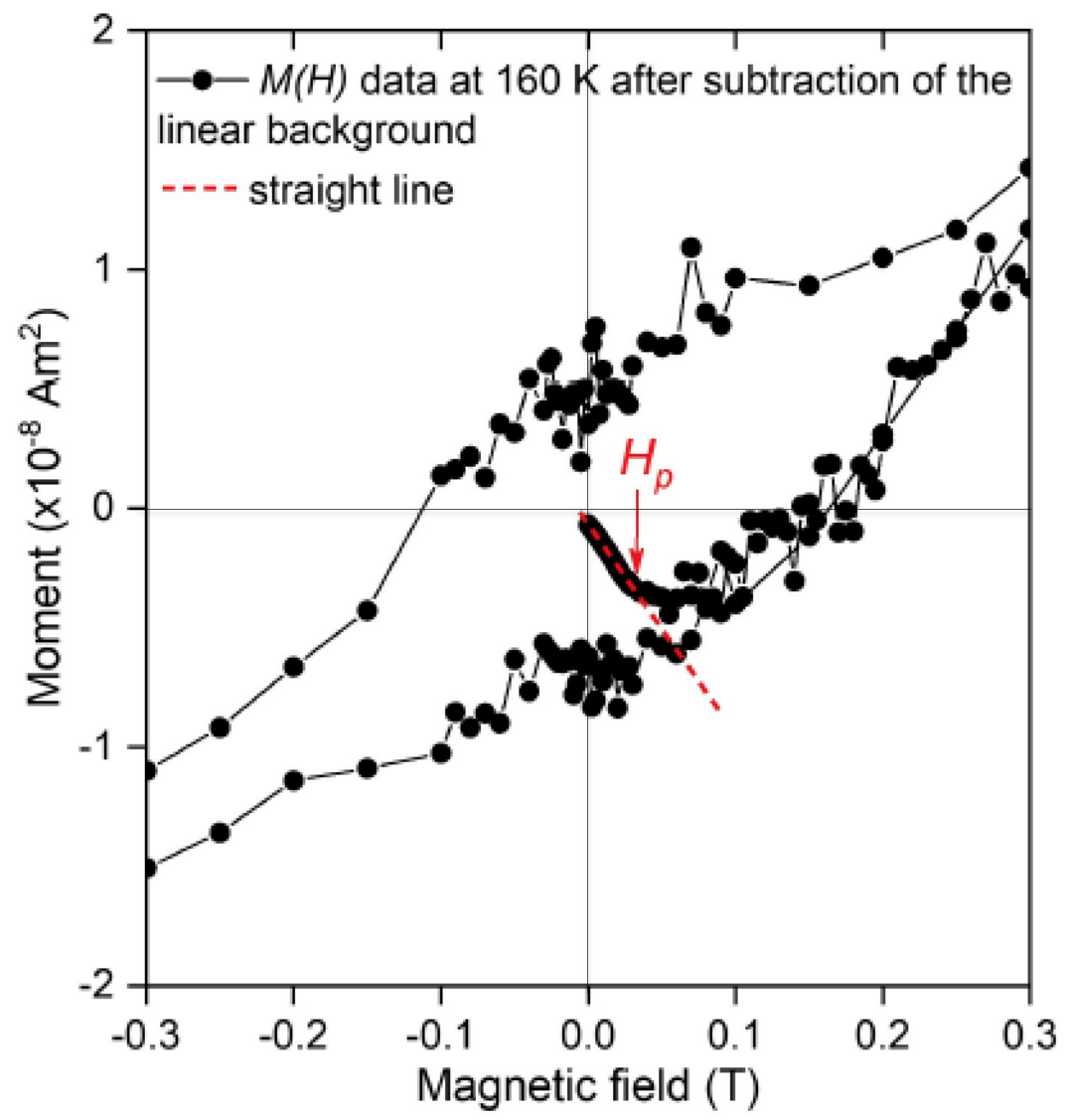}} 
 \caption { Fig. S12c of Ref. \cite{e2021p}, added to the paper when the correction was
 published \cite{correction}.   We have  added the horizontal and vertical  lines going through the origin.
  Note that the moment that resulted after subtraction of the red line in Fig. 2 is positive for applied magnetic field 
  larger than 0.2 T.  If this reflects the behavior of the sample, it implies the sample is paramagnetic for applied fields
  larger than 0.2T.}
 \label{figure1}
 \end{figure} 
 
 In Fig. 4 we show what results from subtracting the red line in Fig. 3 from the measured data, which as
 explained by the authors \cite{correction} was the procedure used to obtain the data in Fig. 1. According to Minkov et al \cite{e2021p},  the results shown in Fig. 4 reflect the 
 behavior of the $H_3S$ sample under applied magnetic field after background subtraction, and the deviation of the virgin curve from the  linear behavior
 indicated by the dashed red line at magnetic field approximately 30mT   (seen more clearly in Fig. 1, curve labeled rd) indicates that the magnetic
 field starts to penetrate the sample at that field, hence is labeled $H_p$. According to the authors, this reflects the lower
 critical field for the superconducting sample, after correcting for a demagnetization factor estimated to be 8.5
 in Ref. \cite{e2021p}.

         \begin{figure} [t]
 \resizebox{8.5cm}{!}{\includegraphics[width=6cm]{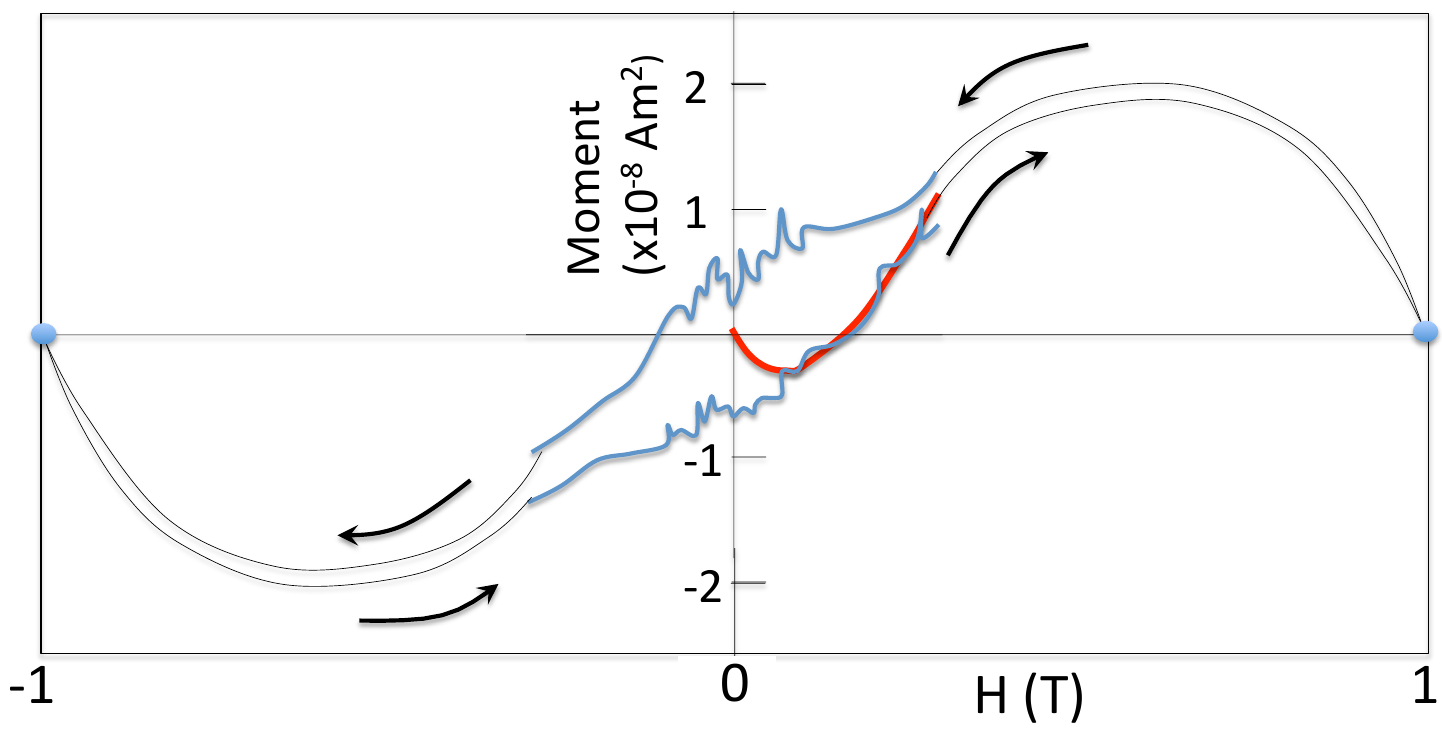}} 
 \caption { Educated guess of what the moment inferred after subtraction of the straight red line from the
 measured data look like in the range of magnetic field up to 1T. Note that the curves go to zero at 1T. }
 \label{figure1}
 \end{figure}

 Note however that the largest diamagnetic moment magnitude in the virgin curve shown in Fig. 4, at field approximately 90 mT, is approximately
 $0.3\times 10^{-8} Am^2$, which is more than 6 times smaller than the negative moment for zero field that was measured shown 
in  Fig. 3, approximately $-2\times 10^{-8} Am^2$, which clearly does not reflect the diamagnetic 
 properties of a hydride superconductor sample. Note also that the total diamagnetic signal measured  at that field was 
 approximately $-6\times 10^{-8} Am^2$, i.e.   more than 20 times larger in magnitude than the diamagnetic moment attributed to the sample.

Also note that the magnetic moment shown in Fig. 4 turns positive for applied magnetic field larger than 0.2T. 
 If indeed it reflects the properties of the $H_3S$ sample,  as interpreted by Minkov et al \cite{e2021p}, it means that the sample is paramagnetic for 
 applied fields larger than 0.2T. This is not consistent with the fact that this is supposed to be 
 a type II superconductor, with an upper critical field   estimated to be
97T \cite{e2021p}.

 Next, we would like to know the behavior of the magnetic moment for  applied field larger than 0.3T, up to 1T.
 Unfortunately it is impossible to discern it from the published data, left panel of Fig. 2, due to the extremely low
 resolution of the figure. Fortunately, we know one point: for magnetic field 1T, the red line shown on the left panel
 of Fig. 3 goes through the measured data point at 1T. That is how the red line was constructed according to
 Ref. \cite{correction}. This implies that the moment extracted from subtraction of the red line, shown in 
 Fig. 4, has to go to zero at magnetic field 1T. In Fig. 5 we show our educated guess of what the 
 data may   look like: we continued the curves smoothly, matching the first derivative at the largest field where we
 had information, and making it reach smoothly the blue known point of zero moment at H=1T.
 
%            \begin{figure} [t]
% \resizebox{8.5cm}{!}{\includegraphics[width=6cm]{fig6.pdf}} 
% \caption { Educated guess of what the magnetic moment may look like for applied field larger than 1T also. }
% \label{figure1}
% \end{figure} 
% 

            \begin{figure} [t]
 \resizebox{8.5cm}{!}{\includegraphics[width=6cm]{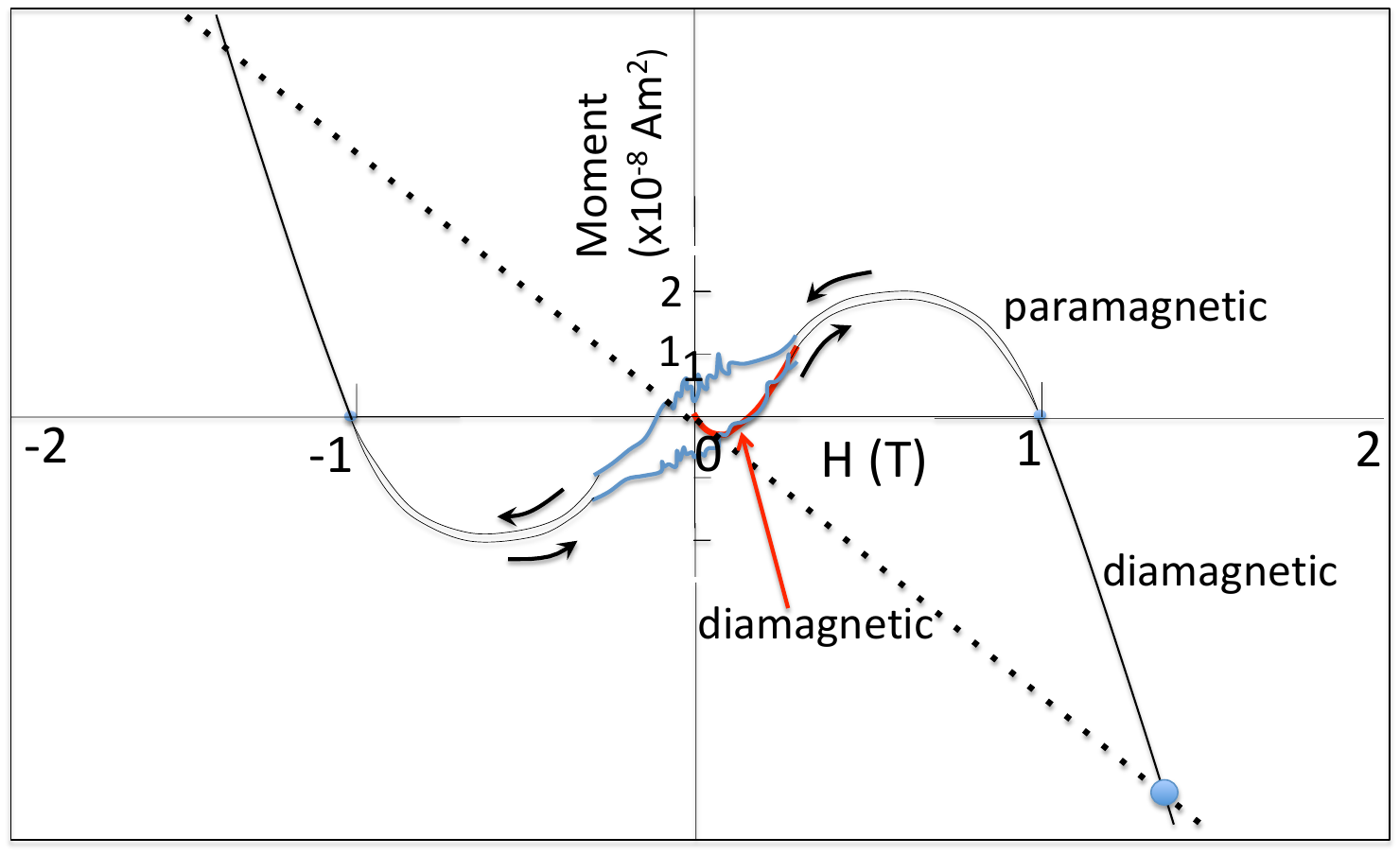}} 
 \caption { Educated guess of what the magnetic moment may look like for applied field larger than 1T.   We have also added the dotted line goes through the origin and through the value of the magnetic moment for $H=1.38T$, and is on top
 of the moment curve (in red) for small fields. For its significance, see text.}
 \label{figure1}
 \end{figure}

%           \begin{figure} [t]
% \resizebox{8.5cm}{!}{\includegraphics[width=6cm]{fig7.pdf}} 
% \caption {Another educated guess of  what the magnetic moment may look like for applied field larger than 1T. 
% The dotted line goes through the origin and through the value of the magnetic moment for $H=1.38T$, and is on top
% of the moment curve (in red) for small fields. For its significance, see text. }
% \label{figure1}
% \end{figure} 
% 

 It would also be nice to know how the sample reacts to magnetic fields larger than 1T. Note for example that
 for the experiments on trapped flux reported by Minkov et al. in Ref. \cite{etrappedp}, magnetic fields
 up to 6T were used. In Fig. 6 we show an educated guess of what the magnetic moment could look like,
 where we have continued the curve beyond 1T with approximately the same slope. Note that it is likely
 that the magnetic moment would change sign again for fields larger than 1T, since it is unlikely that the 
 curve would reach a minimum exactly at 1T, or discontinuously change the sign of its derivative at 1T. So 
 if such behavior reflects the behavior of the sample, it has the remarkable property of being diamagnetic for
 fields smaller than 0.2T, paramagnetic for fields between 0.2T and 1T, and diamagnetic again for fields
 larger than 1T, which is still well below what is expected to be the upper critical field for this material.
 
 If the behavior shown in Fig. 6  was  what was measured,
it   allows us to make the following interesting suggestion. Suppose the authors had decided to draw the
red line shown in  Fig. 3 not according to the criterion they used, namely
{\it ``we have subtracted a linear background from the measured M(H) magnetization data. This
linear background was determined as the straight line connecting two endpoints: the magnetic
moment value at H = 0 T (the starting point of measurements) and the magnetic moment value
at H= 1 T''}, but replacing $H=1T$ by $H=1.38T$. That is the value of H for  the blue point in Fig. 6 where 
our hypothesized curve crosses the dotted line that
goes through the origin. Note that the moment for small fields, indicated by the red curve, that is negative for
field smaller than 0.2T according to the construction of the authors, lies right on top of the dotted line in 
Fig. 6. This means that if the authors had chosen to draw their red line through the $H=1.38T$ data point rather than the one at
1T, upon subtraction of the background the inferred moment for the sample would have been always positive or zero,
i.e. it would indicate that the sample is paramagnetic for all fields.

\section{measured data for other temperatures}
Figure 1, which is Fig. 3a of Ref. \cite{e2021p}, shows curves for 9 different temperatures. Only for one temperature, T=160K, were the raw data shown in the new
Figure S12 published with the correction on September 1, 2023 \cite{correction}. For three other curves in Fig. 1, for temperatures 180K, 140K, and 100K,
hysteresis loops were shown in Fig. S10 of \cite{e2021p}. However, it is hard to draw conclusions from them because they presumably  contain undisclosed shifts of the origin, such as the one
between   Fig. 2 and Fig. 3 here, and in addition the red line used for subtraction similarly to the red line shown 
in Fig. 3  has not been disclosed for these temperatures. For the remaining 5 temperatures, labeled with question
marks in Fig. 1, namely 120K, 80K, 60K, 40K and 20K, neither the origin shifts nor the  hysteresis loops measured nor the red line used
in the subtraction process that was performed to arrive at the data shown in Fig. 1, have been disclosed.

Given this, the relation between the published curves shown in Fig. 1 and what was measured remains largely  unknown.
As a consequence, it is impossible to ascertain whether or not the curves shown in Fig. 1 contain any relevant
information about the physical sample.
 
\section{other transformations performed on measured data}

The subtraction of a linear function and the shift of the origin  were not the only transformations that were performed in going from the 
measured data to the published data that originally had been reported as measured \cite{original}.
According to Ref. \cite{correction},
{\it ``the data were normalized to H = 15 mT data so that to
have the same initial linear M(H) slope.''}, and 
{\it ``we performed additional linear transformations so that the curves have the same
initial linear M(H) slope. Importantly, these linear manipulations do not affect the onset of the
deviation of the M(H) virgin curve from the linear dependence''}.

Neither the paper nor the Author Correction give any information on the magnitude of these linear transformations
that apparently changed the initial slope of the curves, nor do  they give a reason for why doing such transformations
is justified. If the diamagnetic signal measured is due to superconductivity, one would expect the initial slope to
depend on sample size and geometry but be independent of temperature. Thus, under that assumption one could
perhaps argue that changes in slope are not intrinsic and instead result from experimental artifacts and should be
normalized away. Alternatively, without starting from the assumption that the samples are superconductors, one could infer from
the fact that the slope of the signals changes with temperature the conclusion that the signals do not
originate in superconductivity.

\section{unavailability of underlying data}
The published paper Ref. \cite{e2021p} contains a data availability statement, that reads:
{\it``The data that support the findings of this study are available from the corresponding
authors upon reasonable request.''} Clearly, the measured data before background subtraction for the 9 curves shown
in Fig. 3a of Ref. \cite{e2021p} for $H_3S$, and for the 4 curves shown in Fig. 3b for $LaH_{10}$, are data that support the findings of
that study. We now know that those data were processed by shifting the origin, subtracting
a very large linear diamagnetic background, and performing additional normalizations, but do not know
any of the details of these procedures.

Clearly, requesting the measured data to understand the relation between what was measured and what was published, and to what extent the transformations performed may affect the interpretation of the 
significance of the processed data to the understanding of the physical properties of the samples under
study, is
a reasonable request. Yet, the measured data are not available. We have  requested these data repeatedly  from the authors
 and from the journal starting on January 11, 2023, and not received any of them to date  (September 5, 2023)
 \cite{datajournal,datamp}.

\section{on the nature of the ``Author Correction''}
The Author Correction \cite{correction}  states that the original manuscript \cite{original}  contained errors, namely the omission of the information that the
presented data had undergone various {\it ``linear manipulations''}. If those errors had been inadvertent, these transformations
and background subtraction should have been acknowledged immediately when the inconsistency in the published data
was called to the attention of the authors in October 2022 \cite{emails}, and a correction should have been published long ago.
The fact that this didn't happen suggests that the ``errors'' were not inadvertent but a deliberate attempt to
hide information. This is supported by the
fact that to this date the authors refuse to make their underlying data available for examination by readers  \cite{datajournal,datamp}.

What could be the possible reasons for why this is done? One possibility is that the {\it ``manipulations''} \cite{correction}  of the data
involved steps that are incompatible with generally accepted scientific norms, and this would be potentially revealed if
the underlying data were available  to readers.

\section{incompatibility of published raw data with published processed data}

The caption of Fig. S12  of Ref. \cite{e2021p}, that was added to the paper when the Author Correction \cite{correction} 
was published, reads:
{\it ``Subtraction of the linear background for the better illustration of the value of $H_p$, at which an
applied magnetic field starts to penetrate into the sample. a, b Raw M(H) magnetization data
measured at T = 160 K (black circles) and the linear background, which was determined as the
straight line connecting two endpoints: the magnetic moment value at H = 0 T (the starting point
of measurements) and the magnetic moment value at H = 1 T (the highest value of the applied
magnetic field). c Corrected M(H) magnetization data after subtraction of the linear background
(black circles). The value of $H_p$, at which an applied magnetic field starts to penetrate into the
sample, was determined as the onset of the evident deviation of the M(H) data from the linear
dependence (red dashed straight line)''}.

This is supposed to explain how the data published in Fig. 3a of Ref. \cite{e2021p}  (Fig. 1 here) 
for T=160K were obtained.
So we expected that the data points in Fig. S12c of the corrected Ref. \cite{e2021p} (Fig. 4 here), obtained after the background subtraction, would be the same as
the data points in Fig. 3a. This is however not the case. Fig. 7 shows the superposition of the two sets
of data. The black dots show considerable more scatter than the red dots. Presumably, the black dots
include both the data from the virgin curve and from the return curve in the hysteresis cycle, as seen
in Fig. 4. However, there is clearly no correspondence between black and red points. 
Starting at field 30mT, there are 16 red points from Fig. 1  that lie on a smooth curve.
There is no subset of black points that could correspond to those red points.

One could speculate that the black points and the red points originated in different runs. However
there is much larger scatter in the black points than in the red points. In addition, the purpose
of Fig. S12 is to illustrate how the data in Fig. 3a of the paper were obtained, for which one would
expect that a subset of the black points in Fig. 8 would match all the red points.

We conclude that the relation between measured data and the data published in Fig. 3a of
Ref. \cite{e2021p} remains completely obscure even after the Author Correction,
even for the single example addressed in the Correction, namely the inclusion of 
Fig. S12 to the manuscript.  

It is also very peculiar that the authors chose to illustrate their data manipulation with 
the curve for 160K, instead of with one of the lower temperature curves in Fig. 1 
where the features supposedly associated with superconductivity are much sharper, so that
one would expect to see their signature in the raw data before background subtraction 
to be much more prominent than for the high temperature curve chosen.

           \begin{figure} [t]
 \resizebox{8.5cm}{!}{\includegraphics[width=6cm]{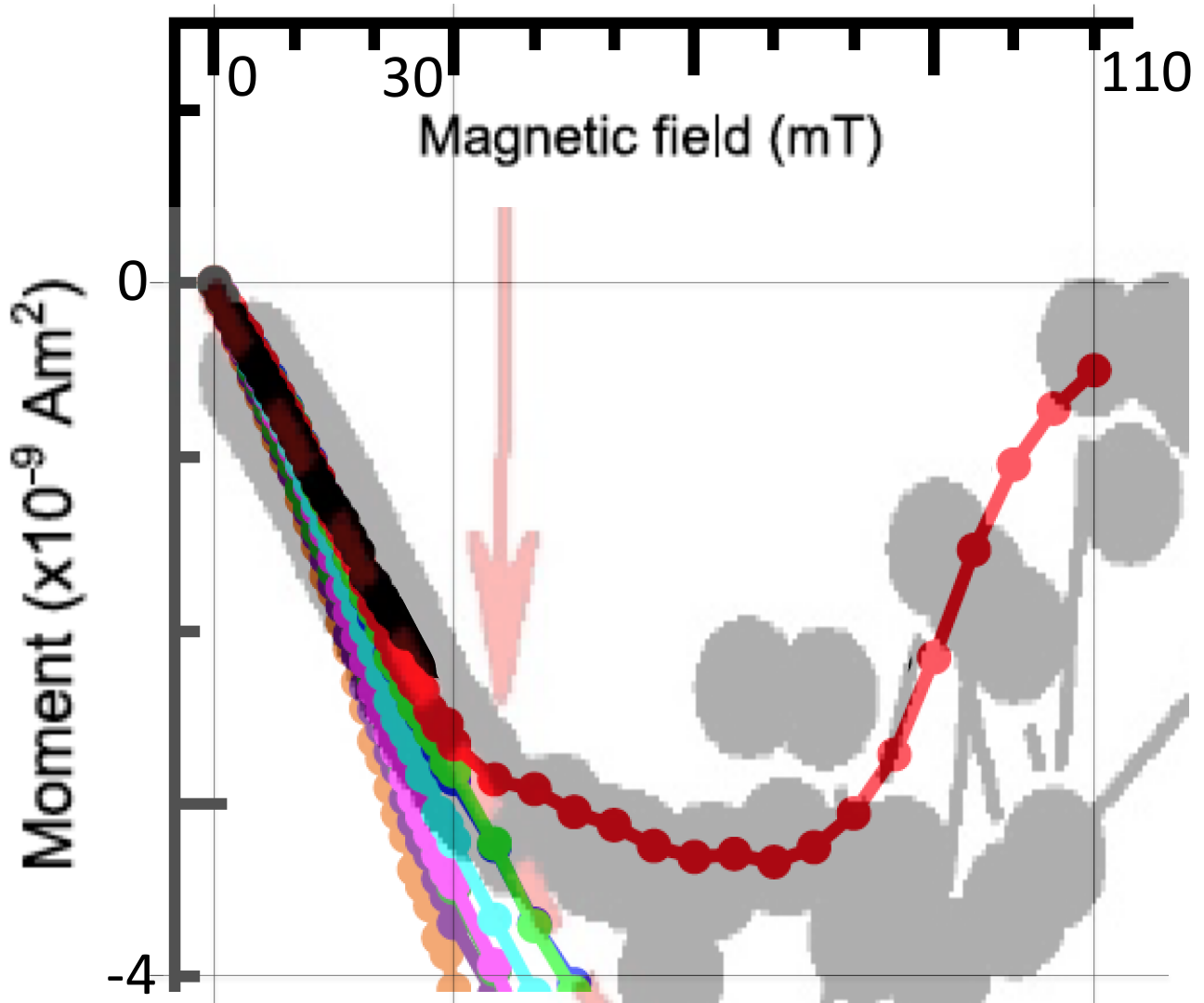}} 
 \caption {Published data from  Fig. 3a of Ref. \cite{e2021p} (Fig. 1 here) for 160 K (red dots), superposed to data from   Fig. S12c of 
 Ref. \cite{e2021p} (Fig. 4 here) (black dots). Thin vertical lines indicate field values 0mT, 30mT and 110mT.
 Thin horizontal lines indicate moment $0\times 10^{-9}Am^2$ and $-4\times 10^{-9}Am^2$. }
 \label{figure1}
 \end{figure} 

\section{implications for trapped flux experiments}
In Ref. \cite{etrappedp}, results of experiments were reported where the same samples of the same materials discussed here
were cooled to low temperature (10K), then a magnetic field of varying magnitude up to 6T was applied, then after 1 hour it was removed,
and then the remnant magnetic field was measured. It was argued that the remnant field was trapped flux, showing 
evidence that supercurrents circulate in these materials and that these materials are strong superconductors
with very strong pinning centers, and the results were interpreted using the Bean model for hard superconductors.
It was deduced that for $H_3S$ the magnetic field $H^*$ where the applied field reaches the center of the sample
has magnitude $H^*\sim 0.8T$. 

We do not know the raw magnetization data before background subtraction and other transformations
 that were measured for low temperature for Ref. \cite{e2021p}, e.g. for the curve
labeled 20K in Fig. 1. We can however guess that the inferred magnetization of the sample after background 
subtraction will look qualitatively similar to Figs. 4 and 5. However, one would expect
for a superconductor with estimated upper critical field 97T \cite{e2021p} as $H_3S$ is,   that
necessarily the diamagnetic response should persist for fields $H^*=0.8T$ and much larger.
Even for an ideal type II superconductor without pinning centers the diamagnetic response persists up to the
upper critical field, and for superconductors with pinning centers the diamagnetic response should be even
stronger since vortices that would weaken the diamagnetic response are prevented from penetrating due to
pinning centers. Therefore the behavior shown in Figs. 4 and 5, and as pointed out in Ref. \cite{hmscreening} even the behavior shown in Fig. 1 with the
magnetization even at low temperatures being strongly reduced already  for $H\sim 200 mT$, is   incompatible with superconducting behavior.
Consequently, if the hysteresis loop shown in Fig. 4 is not reflecting the magnetization of a superconducting sample,
there is no reason to believe that the magnetic moment that remains after the field is first applied and
then removed, i.e. the value of the moment at the point where the upper curve of Fig. 4 crosses
the y-axis,  
originates in superconducting currents creating trapped flux, as claimed in Ref. \cite{etrappedp}.

We have also pointed out \cite{hmtrapped}  that the field dependence of the trapped flux reported by Minkov et al. under zero field cooling,
which is linear for  small fields  \cite{etrappedp}, is incompatible with the conclusion that it originates in superconductivity,
since it would be quadratic in that  case, as seen for example in similar recent experiments on $MgB_2$
samples \cite{budko}.

\section{discussion}
From what we have now learned from Ref. \cite{correction},
the experimental apparatus used to perform the measurements reported in Ref. \cite{e2021p} has a
{\it ``significant diamagnetic response''}, for reasons that have not been explained. The magnitude of that response is several times larger
that the magnitude of the diamagnetic moments attributed to the sample in Ref. \cite{e2021p}. The field dependence inferred for the magnetic
moment of the sample after background subtraction is totally incompatible with the behavior of a superconducting
sample, as shown in Figs. 4-6. All of this indicates that the observed diamagnetism   is not due to superconductivity of the sample.

The authors  of Ref. \cite{e2021p} seem to believe that the hysteresis shown in their magnetization cycles for the raw data is clear evidence for
superconductivity. However there could be a variety of reasons for this hysteresis due to the experimental apparatus that 
have nothing to do with superconductivity. One obvious check would be to perform these experiments with a 
sample before the laser treatment that is supposed to render it superconducting, and show that in that case
no hysteresis is seen. The fact that the authors have not reported this simple check
invalidates their claim that their presented results are evidence for superconductivity in their samples.

We have also argued in this paper that their background subtraction procedure that was used is arbitrary, and showed an
example where  choosing a different equally plausible straight line to subtract from hypothesized measured   raw data would render a signal that is always paramagnetic. We have been unable to judge the validity of the authors'  other additional normalization procedures performed
because they have not disclosed the details of them nor have they presented  arguments for why such procedures
would be justified.

Finally, the fact that the  underlying data are not made available  \cite{datajournal,datamp}, and particularly not for low temperatures
where signatures of superconductivity should be more apparent,
is a big red flag that suggests that other features of the measured data may not be compatible with superconductivity.
Moreover, the fact that underlying data are not made available  \cite{datajournal,datamp} leaves open the possiblity
that some of the manipulations of the measured data that were  performed to arrive at the published data may not be
compatible with accepted scientific practice.

Eight years after the reported experimental discovery of high temperature superconductivity in a hydride under high pressure \cite{e2015},
there is no convincing magnetic evidence that these materials are superconductors \cite{pers}. The analysis
in this paper eliminates the strongest claims in that regard. The resistance evidence
has also been called into question \cite{hmnature,dc,nonstandard,hamlin,enormous,vdp}. Readers should draw their own conclusions.

%\begin{acknowledgments}
%
%\end{acknowledgments}


\begin{references}
     
        \bibitem{e2021p} V. S. Minkov et al, ``Magnetic field screening in hydrogen-rich high-temperature superconductors'',
\href{https://www.nature.com/articles/s41467-022-30782-x} {Nat Commun 13, 3194 (2022)}.

\bibitem{correction} V. S. Minkov et al, ``Author Correction: Magnetic field screening in hydrogen-rich high-temperature superconductors'',
\href{https://www.nature.com/articles/s41467-023-40837-2}{Nat Commun 14, 5322 (2023)}.

\bibitem{etrappedp}     V. S.  Minkov et al,
``Magnetic flux trapping in hydrogen-rich high-temperature superconductors'',
\href{https://www.nature.com/articles/s41567-023-02089-1}{Nat. Phys. {\bf 19}, 1293 (2023)}.



\bibitem{review} Mikhail Eremet, Maddury S. Somayazulu, Artem R. Oganov and Ioulia A. Ovchenkova, 
``Phenomena of hydrides'',
\href{https://pubs.aip.org/aip/jap/article/132/18/180401/2837665}{Journal of Applied Physics 132, 180401 (2022)}
and references therein.

     \bibitem{pers}
     J. E. Hirsch, ``Are hydrides under high pressure high temperature superconductors?'',
     \href{https://doi.org/10.1093/nsr/nwad174}{National Science Review, nwad174 (2023)}
and references therein.

  \bibitem{pickett}  W. E. Pickett, ``Room temperature superconductivity: The roles of theory and materials design'',
    \href{https://journals.aps.org/rmp/abstract/10.1103/RevModPhys.95.021001}{Rev. Mod. Phys. 95, 021001 (2023)}.

\bibitem{zurek} K. P. Hilleke and E. Zurek,
``Rational Design of Superconducting Metal Hydrides via Chemical Pressure Tuning'',
\href{https://onlinelibrary.wiley.com/doi/abs/10.1002/anie.202207589}{Angewandte Chemie  61, e202207589  (2022)}.




\bibitem{original} The original version of Nat Commun 13, 3194 (2022), which was 
on-line between 06/09/2022 and 08/31/2023,
is no longer available at the journal's website but can be found at 
\href{https://jorge.physics.ucsd.edu/e2021p.pdf}{https://jorge.physics.ucsd.edu/e2021p.pdf}.

\bibitem{emails} First email from JEH to authors of Ref. \cite{e2021p} asking for clarification of magnetization data was on 
October 18, 2022, 10th email was on May 10, 2023. No clarification was provided.

\bibitem{hmscreening}  J. E. Hirsch and F. Marsiglio, ``On Magnetic Field Screening and Expulsion in Hydride Superconductors'',
\href{https://link.springer.com/article/10.1007/s10948-023-06569-6}{J Supercond Nov Magn 2023;  36: 1257–1261}.

\bibitem{datajournal} We were informed by a Nature Comm. editor  that the editors are in possession of these data,
 however they consider them ``confidential'' following the request of the authors of
 Ref. \cite{e2021p} and for that reason cannot be shared with us.
 The same editor told us that the journal will not publish an Editor Note informing readers that there are 
 restrictions on data availability for Ref. \cite{e2021p}   because the data   are available to the editors.
 
 \bibitem{datamp}The director of the Institute where the measured data for the research reported in Ref. \cite{e2021p} were obtained
 was informed by one of us (JEH) 5 months ago  (April 2, 2023) that the measured data are not being shared.





%\bibitem{editor} The editor is Dr Prabhjot Saini, Chief Editor, Physics and Earth Sciences, 
%Nature Communications.

\bibitem{hmtrapped} J. E. Hirsch and F. Marsiglio,
``Evidence Against Superconductivity in Flux Trapping Experiments on Hydrides Under High Pressure'', \href{https://link.springer.com/article/10.1007/s10948-022-06365-8}
{J Supercond Nov Magn 2022; 35: 3141–3145}.

\bibitem{budko}
V.  Ksenofontov, 
``Magnetic flux trapping in high-$T_c$ superconducting
hydrides'', in \href{https://www.superstripes.net/conferences/superstripes-2023}{Superstripes 2023},
June 28, 2023, 11-th slide, S. L. Budko et al, submitted.



          \bibitem{e2015} A.P. Drozdov, M.I. Eremets, I. A.Troyan, V. Ksenofontov  and S. I. Shylin,
     `Conventional superconductivity at 203 kelvin at high pressures in the sulfur hydride system',
     \href{https://www.nature.com/articles/nature14964}{Nature 525, 73-76 (2015)}.
     




 
 \bibitem{hmnature} J. E. Hirsch and F. Marsiglio, ``Unusual width of the superconducting transition in a hydride'',
\href{https://www.nature.com/articles/s41586-021-03595-z}{Nature 596, E9  (2021)}.

  \bibitem{dc} M. Dogan and M.  L. Cohen, ``Anomalous behavior in high-pressure carbonaceous sulfur hydride'', 
\href{https://www.sciencedirect.com/science/article/pii/S0921453421000344}{Physica C 583, 1353851 (2021)}.

\bibitem{nonstandard}
J. E. Hirsch and F. Marsiglio, ``Nonstandard superconductivity or no superconductivity in hydrides under high pressure'', \href{https://journals.aps.org/prb/abstract/10.1103/PhysRevB.103.134505}{Phys. Rev. B 103, 134505 (2021)}.
 

      \bibitem{hamlin} J. J. Hamlin, ``Vector graphics extraction and analysis of electrical resistance data in Nature volume 586, pages 373-377 (2020)'',
\href{https://arxiv.org/abs/2210.10766}{arXiv:2210.10766 (2022)}.



 \bibitem{enormous}
        J. E. Hirsch,
       ``Enormous variation in homogeneity and other anomalous features of room temperature superconductor samples'',
       \href{https://link.springer.com/article/10.1007/s10948-023-06593-6}{ J Supercond Nov Magn 36, 1489 (2023)}.








\bibitem{vdp}  J. E. Hirsch, ``Electrical resistance of hydrides under high pressure: evidence of superconductivity or confirmation bias?'', \href{https://link.springer.com/article/10.1007/s10948-023-06594-5}{  J Supercond Nov Magn 36, 1495 (2023)}.

     
     
            \end{references}
 \end{document}